\documentclass[11pt,preprint]{aastex}

\usepackage{graphics,graphicx}
\usepackage{natbib}
\usepackage{epstopdf}
\newcommand{\gtorder}{\mathrel{\raise.3ex\hbox{$>$}\mkern-14mu
            \lower0.6ex\hbox{$\sim$}}}
\newcommand{\ltorder}{\mathrel{\raise.3ex\hbox{$<$}\mkern-14mu
            \lower0.6ex\hbox{$\sim$}}}

\shorttitle{A Wind Accretion Model for HLX--1}
\shortauthors{}

\begin{document}

\title{A Wind Accretion Model for HLX--1}

\author{M. Coleman Miller\altaffilmark{1}, Sean A. Farrell\altaffilmark{2}, and Thomas J. Maccarone}
\altaffiltext{1}{Department of Astronomy and Joint Space-Science Institute, University of Maryland, College Park, MD 20742--2421, USA}
\altaffiltext{2}{Sydney Institute for Astronomy (SIfA), School of Physics, The University of Sydney, NSW 2006, Australia}
\altaffiltext{3}{Department of Physics, Texas Tech University, Lubbock TX, 79409, USA}

\begin{abstract}

The brightest ultraluminous X-ray source currently known, HLX--1, has been observed to undergo five outburst cycles.  The periodicity of these outbursts, and their high inferred maximum accretion rates of $\sim{\rm few}\times 10^{-4}~M_\odot~{\rm yr}^{-1}$, naturally suggest Roche lobe overflow at the pericenter of an eccentric orbit.  It is, however, difficult for the Roche lobe overflow model to explain the apparent trend of decreasing decay times over the different outbursts while the integrated luminosity also drops.  Thus if the trend is real rather than simply being a reflection of the complex physics of accretion disks, a different scenario may be necessary.  

We present a speculative model in which, within the last decade, a high-mass giant star had most of its envelope tidally stripped by the $\sim 10^{4-5}~M_\odot$ black hole in HLX--1, and the remaining core plus low-mass hydrogen envelope now feeds the hole with a strong wind.  This model can explain the short decay time of the disk, and could explain the fast decrease in decay time if the wind speed increases with time.  A key prediction of this model is that there will be excess line absorption due to the wind; our analysis does in fact find a flux deficit in the $\sim 0.9-1.1$~keV range that is consistent with predictions, albeit at low significance.  If this idea is correct, we also expect that within tens of years the bound material from the original disruption will return and will make HLX--1 a persistently bright source.

\end{abstract}


\section{Introduction}
\label{sec:intro}

The source 2XMM J011028.1--460421, commonly known as HLX--1, was identified by \citet{2009Natur.460...73F} as being hosted by the z=0.0224 galaxy ESO 243--49.  At the $\sim 93$~Mpc distance of this galaxy, the peak flux of this source corresponds to an isotropic bolometric luminosity of $\sim 2\times 10^{42}~{\rm erg~s}^{-1}$.  The off-center location in the galaxy, the variability of the source, and its spectral properties make this the best current candidate for an intermediate-mass black hole (IMBH).  The mass of the hole has been estimated from spectral fitting, matching of spectral states as a function of Eddington ratio with stellar-mass sources, and a limitation of the luminosity to less than the Eddington luminosity, to be $M\sim 10^{4-5}~M_\odot$ \citep{2011ApJ...734..111D,2011ApJ...743....6S,2012ApJ...752...34G,2012Sci...337..554W}.

The source has now been seen to undergo five outbursts with a period of very close to a year \citep{2013MNRAS.428.1944S}.  Intriguingly, and as we discuss in Section~\ref{sec:obs}, the fifth outburst was delayed by a couple of weeks compared to predictions based on the previously observed period.  Nonetheless, the most natural interpretation of the period is that it is an orbital period.  If so, the factor of $\sim 40$ variation in flux over a cycle implies that the orbit is eccentric \citep{2011ApJ...735...89L,2013MNRAS.428.1944S}.  The high implied peak mass accretion rate $\sim {\rm few}\times 10^{-4}~M_\odot~{\rm yr}^{-1}$ is straightforwardly explained by Roche lobe overflow at the pericenter of the orbit.  

If the system is undergoing Roche lobe overflow, then as we discuss in Section~\ref{sec:Roche} the donor star must be a helium star, because the disk inflow time from the tidal radius for a main sequence star is too long, and from a white dwarf is too short, to explain the observed decay times of a few months.  In such a model, we would expect that the inflow time would remain approximately constant from outburst to outburst, with slightly longer observed decay times for outbursts that have smaller fluence because in the standard theory of geometrically thin disks the viscous time is larger for smaller accretion rates.  

However, fits to the outbursts suggest that the decay times have decreased from $\sim 6$ months to $\sim 2-3$ months (see Section~\ref{sec:obs}) and their fluences have dropped by a similar factor.  It could be that this is not a meaningful trend but is instead the result of complex interactions in the accretion disk.  If instead this is a real effect, then we must consider another scenario for HLX--1.

As an example of an alternate scenario, we explore a model in which the donor star transfers mass via a wind.  The advantages of the model are that it can naturally explain the time scale of decay and the rapid change in that time scale, and that because the donor star need not be close to the tidal radius the dynamical state of the system need not be fine-tuned.  The disadvantage of the wind model is that the required outflow rates of $\sim 10^{-3}~M_\odot~{\rm yr}^{-1}$ are at least an order of magnitude larger than any known wind rate.  As a way around this, we suggest in Section~\ref{sec:wind} that within the past several years a high-mass giant that was already close to its critical luminosity was tidally stripped by the IMBH.  This could push the remaining star to or above its new critical luminosity and lead to very high outflow rates.  The existence of such high-mass giants is plausible given the inferred young massive cluster that hosts HLX--1 \citep{2012ApJ...747L..13F,2014MNRAS.437.1208F}.  In this scenario, the transient nature of the system is the key to the changing fluence and decay time.  

In Section~\ref{sec:obs} we present our data analysis of the outbursts. In Section~\ref{sec:Roche} we discuss the Roche lobe overflow model and in particular the expected time scales for inflow and decay.  In Section~\ref{sec:wind} we explore the wind accretion model and show that when the wind speed is comparable to the orbital speed at pericenter, the net angular momentum of the accreted matter can be low enough that the resulting disk is small.  Thus the decay time of the disk can be in the observed range, and a small increase in the wind speed can reduce the decay time by tens of percent. We also report on analysis of the data from this source that was motivated by the wind scenario: we expect significant extra narrow-line absorption in the system if the accretion is via a wind rather than through Roche lobe overflow.  Indeed, although high-quality X-ray spectra have shown no evidence for high neutral hydrogen columns, there is some evidence for narrow features at the expected amplitude, albeit at low significance, in the predicted $\sim 0.9-1.1$~keV range.  Thus although there are too many uncertainties to claim clear confirmation of the prediction, this is an encouraging match with our scenario.  In Section~\ref{sec:dynamics} we explore the possible role of stellar dynamics, and demonstrate that for reasonable stellar number densities the system should be unchanged during our few-year observational window.  We present our conclusions in Section~\ref{sec:concl}, where we emphasize that continued monitoring with Swift will be needed to narrow down the possible explanations for this unique source.  In particular, we note that in our disrupted giant scenario we expect the bound material to return within tens of years, after which the source will be persistently bright for centuries.

\section{Observations of HLX--1}
\label{sec:obs}

\begin{figure}[htb]
\begin{center}
\plotone{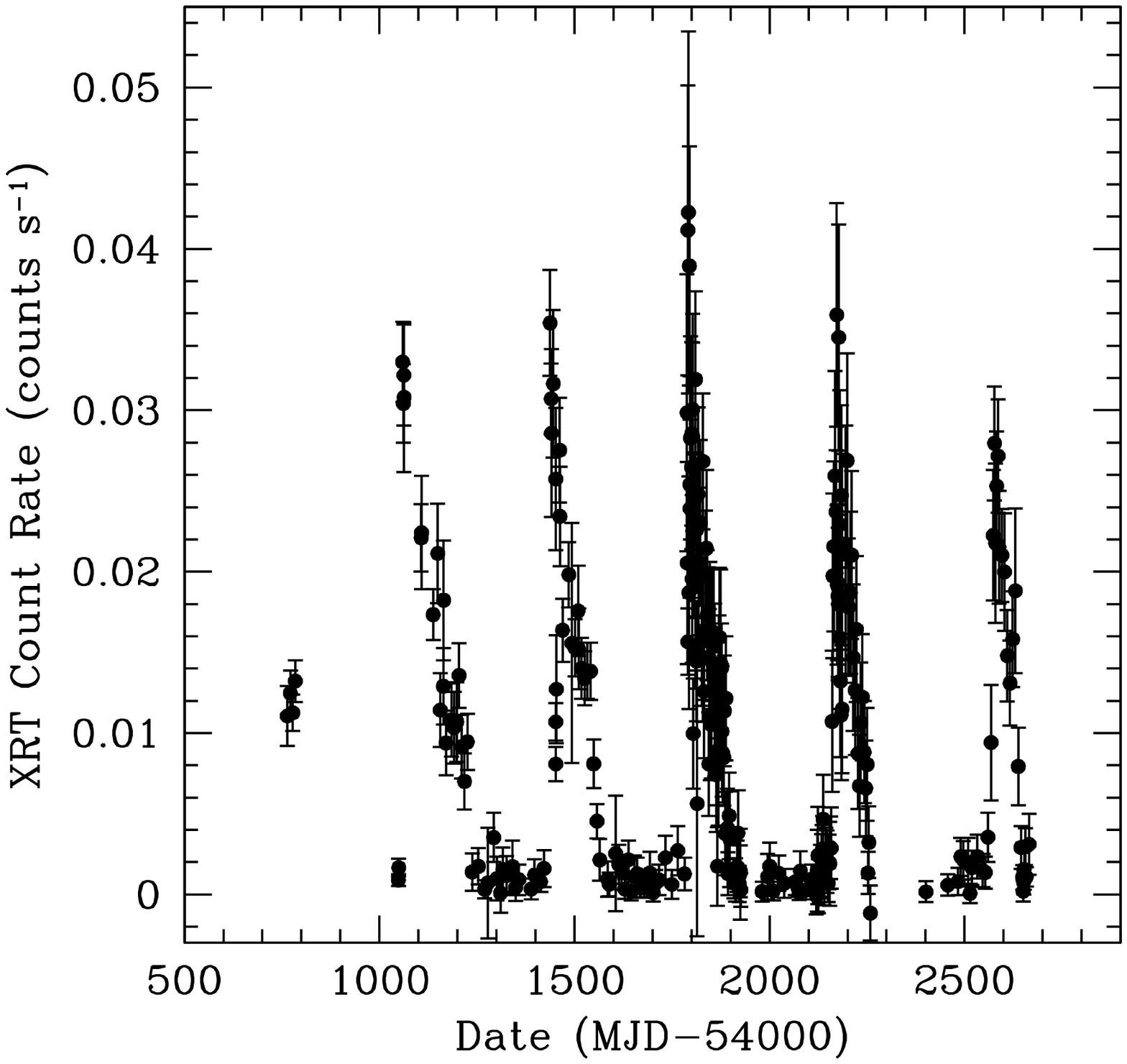}
\caption{\emph{Swift} XRT light curve of the outbursts. Data points taken on the January 13th 2010 (MJD 55520), September 19th 2011 (MJD 55824), and January 17th 2012 (MJD 55944) were removed. The count rates on these dates deviated significantly from the general trend, yet manual inspection of the data found these deviations were spurious, likely due to poor data quality and/or issues with the automatic light curve extraction.}
\label{fig:xrtlc}
\end{center}
\end{figure}

Following its serendipitous discovery with \emph{XMM-Newton} on November 23rd 2004, HLX--1 was observed with the \emph{Swift} X-ray Telescope (XRT) between October 24 and November 13 2008.  Since an initial hiatus of almost 9 months, HLX--1 has been regularly observed with the \emph{Swift} XRT with a monitoring cadence of between 1 day and 3 weeks and a total exposure time of $\sim$600 ks. The XRT light curve, extracted using the online processing facility\footnote{http://www.swift.ac.uk/user\_objects/} \citep{2009MNRAS.397.1177E}, shows the outbursts (see Figure \ref{fig:xrtlc}). Using the epoch folding search technique (the efsearch task in the FTOOLS package), our best estimate of the average recurrence timescale is 372.6 d. We split the light curve into segments covering each of the outbursts (excluding the data taken prior to August 2009) and folded each segment over 372.6 d using the FTOOLS task efold. The folded light curve profiles clearly show a reduction in the integrated {\it Swift} countrate from the first through the fifth outburst. In addition, the time taken to decay from the peak of the outburst to the quiescent level also decreases significantly over time. We summarize the outbursts in Table \ref{tab:outbursts}; here the decay timescale is calculated as the time difference between the peak of the outburst to the first point that is consistent with a rate of 0 counts s$^{-1}$ within the errors. We note that the timescales and peak fluxes for the first two outbursts are likely larger than those quoted, as the cadence of the XRT observations was $\sim$1 -- 3 weeks prior to the beginning of the outbursts, so the precise timing of the peak is unknown.

\begin{table}
\begin{center}
\caption{Outburst parameters.\label{tab:outbursts}}
\begin{tabular}{ccccc}
\tableline\tableline 
Number & Start Date & Peak L$_X$ & Integrated energy & Decay Time\\
 & (MJD) & (10$^{42}$ erg s$^{-1}$) & (10$^{48}$ erg) & (d) \\
\tableline
1 & 55059 & 1.92 & 9.35 & 180\\
2 & 55437 &  1.88 & 8.91 & 130 \\
3 & 55788 & 1.10 & 5.77 & 110\\
4 & 56162 &  0.43 & 3.67 & 90 \\
5 & 56574 & $<$1.5 & $<$6.2 & 75 \\
\tableline
\end{tabular}
\end{center}
\end{table}

Using the average recurrence timescale of 372.6 d we extrapolated backwards from the first outburst to estimate when previous outbursts would have peaked and returned to quiescence. We found that the predicted outbursts were consistent with the observed count rates of the first set of \emph{Swift} XRT observations in November 2008 as well as the original detection in November 2004 with \emph{XMM-Newton}\footnote{\emph{Swift} XRT grade 0--12 count rates were calculated with WebPIMMS using the observed 0.2 -- 12 keV flux and best fit power law spectral model parameters reported in \citet{2011ApJ...743....6S}.}. The field of HLX--1 was also observed on three occasions with \emph{ROSAT} on December 8th 1991 with the PSPCB and on December 27th 1996 and November 23rd 1997 with the HRI. HLX--1 was not detected in any of these observations. Using the pipeline extracted total band images from these data, we estimated 3$\sigma$ count rate upper limits of 0.0012 count s$^{-1}$, 0.0053 count s$^{-1}$, and 0.00076 count s$^{-1}$ for the 1992, 1996, and 1997 observations, respectively. These upper limits were converted into \emph{Swift} XRT grade 0--12 count rates using WebPIMMS assuming a power law spectral model with N$_H$ = 3 $\times$ 10$^{20}$ atoms cm$^{-2}$ and $\Gamma$ = 2.0 \citep[consistent with the best fit to the spectral shape in the third \emph{XMM-Newton} observation when HLX--1 was in the low/hard state;][]{2011ApJ...743....6S}. We thus derived XRT count rates of 0.0008 count s$^{-1}$, 0.01 count s$^{-1}$, and 0.001 count s$^{-1}$ for the 1992, 1996, and 1997 observations, respectively.  If we convert these count rates into luminosities using the model described above, then these limits correspond to luminosities of $7.1\times 10^{40}$~erg~s$^{-1}$, $1.1\times 10^{42}$~erg~s$^{-1}$, and $2.4\times 10^{41}$~erg~s$^{-1}$, respectively.  Comparison with the predicted outbursts found that the observations and upper limits were all consistent with HLX--1 being in the low/hard state and thus undetectable within the \emph{ROSAT} data. However, if the trend towards longer decay times for earlier outbursts applies to the {\it ROSAT} era, the nondetections are surprising. 

\begin{figure}[htb]
\begin{center}
\plotone{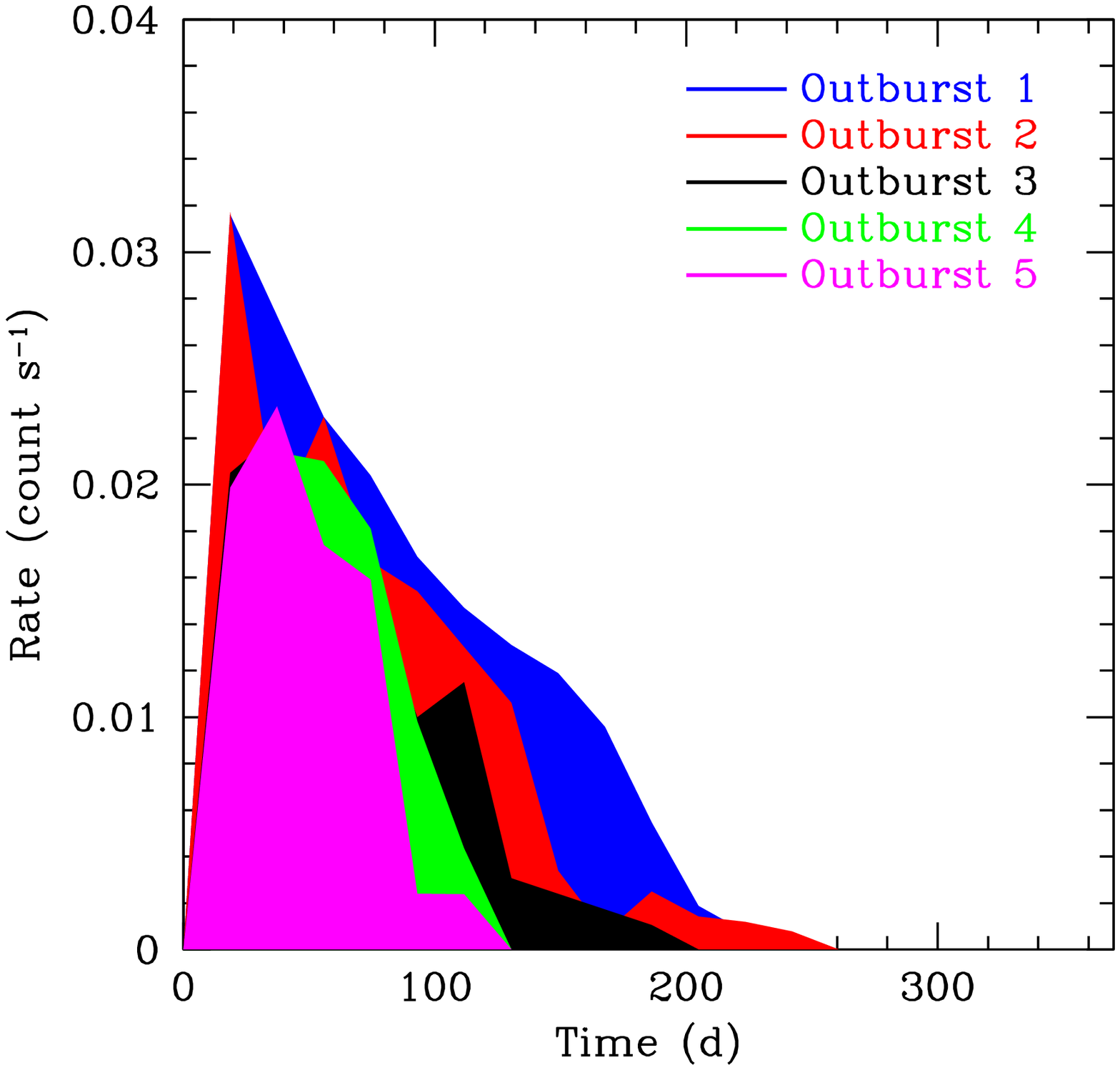}
\caption{\emph{Swift} XRT light curve segments covering the outbursts folded over a period of 372.6 d. The peak flux and decay timescale have clearly decreased over time.  Note that we have aligned the start of the fifth outburst to coincide with the starts of the other outbursts.}
\label{fig:obprofile}
\end{center}
\end{figure}

We next extracted phase resolved spectra for each outburst using the XRT online processing facility in order to derive integrated luminosities.  We extracted spectra containing $\sim$200 counts for outbursts 1, 2, and 3, and spectra containing $\sim$100 counts for outbursts 4 and 5 as the exposure times for the monitoring observations were shorter during the last two outbursts. This resulted in 8, 9, 10, 7, and 7 spectra for each of the outbursts, respectively. We then fitted each spectrum individually in  XSPEC v12.6.0q \citep{1996ASPC..101...17A} with an absorbed irradiated accretion disk model (tbabs*diskir). We used the abundances of \citet{2003ApJ...591.1220L} and froze the column density at 3 $\times$ 10$^{20}$ atoms cm$^{-2}$. We froze the high-energy rollover temperature (kT$_e$) at 100 keV, the fraction of luminosity in the Compton tail (fin) at 0.1, the radius of the Compton illuminated disk (rirr) at 1.2 times the inner disk radius, the fraction of bolometric flux thermalized in the outer disk (fout) at 10$^{-4}$, and the outer disk radius (logrout) at 4. These parameters only affect the spectrum outside the XRT bandpass and therefore cannot be constrained, and were chosen to be consistent with the results of fitting the broad-band spectra of HLX--1 presented in \citet{2014MNRAS.437.1208F}.  We also froze the power law index to 2; this is the average value obtained from all the {\it XMM-Newton} spectral fitting.  We then estimated bolometric de-absorbed luminosities for each phase resolved spectrum using a redshift of 0.0224 \citep{2010ApJ...721L.102W}. Integrated luminosities for each outburst were then estimated using the trapezoidal rule, and are presented in Table \ref{tab:outbursts}. 

\section{Roche lobe overflow model}
\label{sec:Roche}

From Table~\ref{tab:outbursts}, the peak inferred luminosity is $\sim 2\times 10^{42}$~erg~s$^{-1}$.  If the accretion efficiency $\eta\equiv L/({\dot M}c^2)\sim 0.1$, then this luminosity corresponds to an accretion rate of $\sim 3\times 10^{-4}~M_\odot~{\rm yr}^{-1}$.  As discussed by \cite{2011ApJ...735...89L}, the most straightforward way to have a mass transfer rate this high at the pericenter of an eccentric orbit is to have Roche lobe overflow at the pericenter.  \citet{2007ApJ...660.1624S} showed that although the precise distance at which a donor star in an eccentric orbit will overflow its Roche lobe depends on the degree of spin synchronization and other parameters, the modified \citet{1983ApJ...268..368E} formula
\begin{equation}
R_{\rm Roche}=r_p{0.49q^{2/3}\over{0.6q^{2/3}+\ln(1+q^{1/3})}}\; ,
\end{equation}
where $r_p$ is the pericenter distance and $q=m/M$ is the mass ratio of the donor to the IMBH, gives a value for $r_p$ that is accurate to $\sim 20$\%.  If the donor has a mass $m\ll M$, then $q\ll 1$ and $r_p\approx 2q^{-1/3}R_{\rm Roche}$. If the donor is on the zero age main sequence and has $m=m_0M_\odot$ with $m_0\geq 1$, its radius is $R\sim R_\odot m_0^{1/2}$ \citep{1991Ap&SS.181..313D}.  If $M=10^4~M_4 M_\odot$ then because $R_\odot=7\times 10^{10}$~cm, setting $R=R_{\rm Roche}$ implies
\begin{equation}
r_p=3\times 10^{12}~{\rm cm}~m_0^{1/6}M_4^{1/3}\; .
\end{equation}
If the initial angular momentum of the donated matter is the orbital angular momentum of the donor star, then for an initial orbit of eccentricity $e$ this matter will circularize at a radius of $r_p(1+e)$, although additional interactions with the star are likely to truncate the disk.

\begin{figure}[htb]
\begin{center}
\plotone{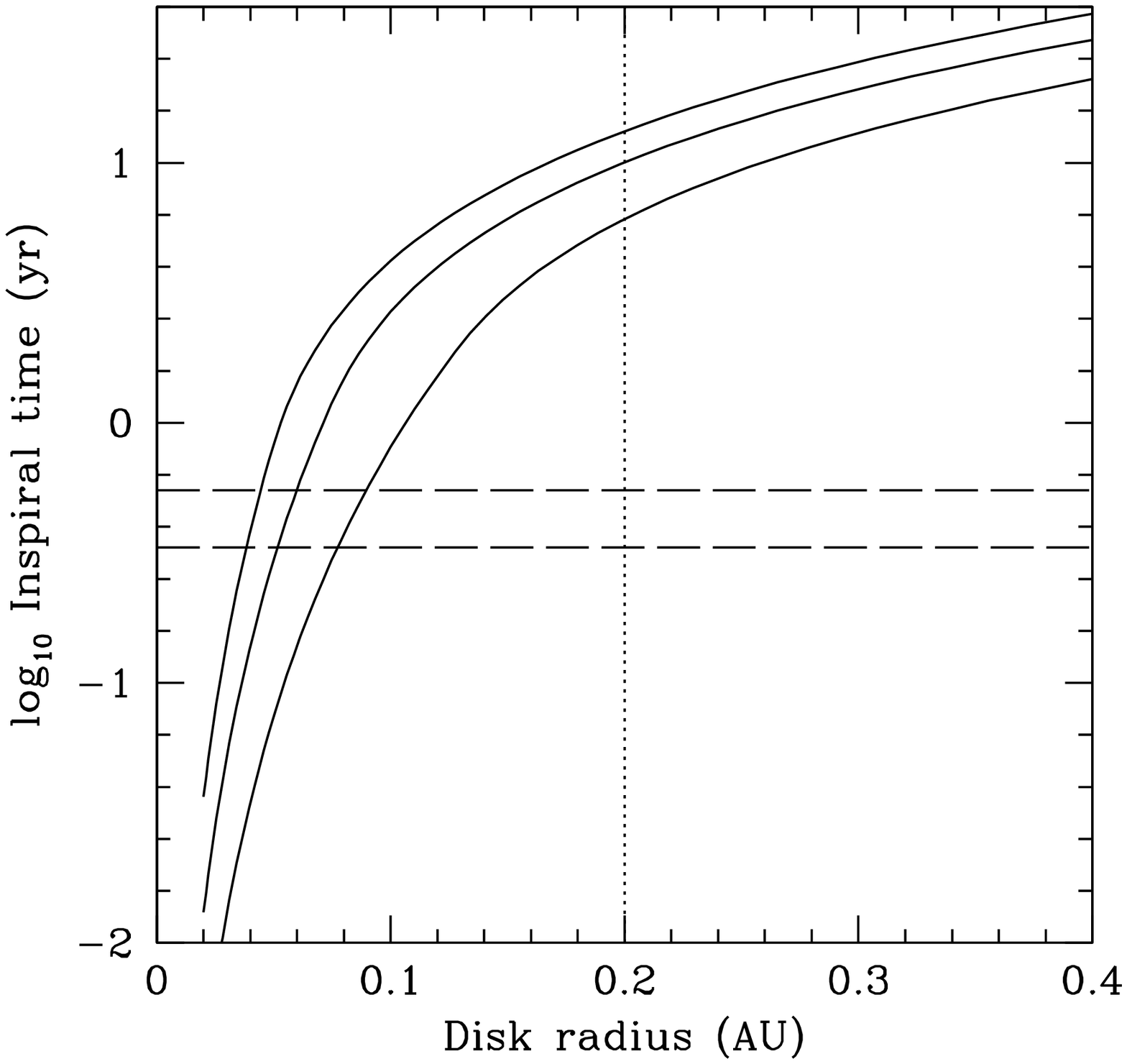}
\caption{Inspiral time from a given disk radius.  Here we assume an IMBH mass $M=10^4~M_\odot$ and consider mass accretion rates (solid curves, from bottom to top) of ${\dot M}=2\times 10^{-4}~M_\odot~{\rm yr}^{-1}$ (comparable to the highest rates inferred for the outbursts), $10^{-4}~M_\odot~{\rm yr}^{-1}$ (a typical average over the outbursts), and $6\times 10^{-5}~M_\odot~{\rm yr}^{-1}$ (average for the fourth outburst). We assume a \citet{1973A&A....24..337S} disk with $\alpha=0.2$.  For such a disk, the inspiral time is dominated by the middle region, where the inward radial speed scales as $v_r\sim \alpha^{4/5}{\dot M}^{2/5}$; the value of $\alpha$ estimated from observations is $\alpha\sim 0.1-0.4$ \citep{2007MNRAS.376.1740K}, so the inflow time could be a factor $\sim 2$ higher or lower than we estimate.  The vertical dotted line shows the radius at which a star of mass $10~M_\odot$ and radius $\sim 3R_\odot=2\times 10^{11}$~cm would donate mass to the IMBH.  The top horizontal dashed line is at the $\sim 6$ month decay time seen in the first outburst, and the bottom horizontal dashed line is at the $\sim 3$ month decay time found for the fourth outburst.  This calculation demonstrates that if the matter spirals in from the radius of Roche lobe overflow, it is difficult to have the short observed decay times.  In addition, because the overflow radius should be approximately constant, it is difficult to change the inspiral time significantly.}
\label{fig:inspiral}
\end{center}
\end{figure}

If we assume that the donated matter forms a \citet{1973A&A....24..337S} disk, we can integrate the inward radial speed to find the characteristic time for the matter to flow from the pericenter to the IMBH.  This is the time scale on which the matter deposited at the pericenter will drain into the hole, so it should be compared with the observed decay time scales of a few months.  Figure~\ref{fig:inspiral} shows the predicted decay times for accretion rates of $2\times 10^{-4}~M_\odot~{\rm yr}^{-1}$, $10^{-4}~M_\odot~{\rm yr}^{-1}$, and $6\times 10^{-5}~M_\odot~{\rm yr}^{-1}$ (characteristic of the highest inferred accretion rate, the average peak rates for all the outbursts, and the minimum peak rate) and a Shakura-Sunyaev parameter $\alpha=0.2$.  Note that the much shorter inspiral time estimated by \citet{2013MNRAS.428.1944S} comes from his assumption that the disk aspect ratio is $h/r=0.1$; this is far larger than is expected from a Shakura-Sunyaev disk of this size.  Of course, if the donor has evolved off the main sequence to larger radii, the pericenter distance and thus the time for Roche lobe overflow would be significantly greater than if the donor is on the main sequence.  In addition, because the inspiral time is slightly longer for smaller accretion rates, one would expect that if the accretion rate drops (as it appears to in the later cycles) the decay time would increase slightly, rather than decreasing by a factor of $\sim 2$ as in the observations.  Thus although the overall time scale might be matched if a lower-mass main sequence star were the donor, it would still be expected that the decay time would increase slightly as the accretion rate declines.  As noted by \citet{2011ApJ...735...89L}, the surge of matter associated with Roche lobe overflow at the pericenter might produce rapid transient accretion, so the fast rises of the outburst light curves are not necessarily problematic for the Roche lobe picture.  The difficulty comes in the longer-term evolution of the disk.

There is also a significant amount of fine-tuning involved in having the donor star come close enough to donate mass but not so close that it is tidally disrupted.  For a star such as a main sequence star, which does not have a distinct core-envelope structure, to donate $\sim 10^{-5}-10^{-4}~M_\odot$ per orbit means that the outer few scale heights are being stripped off in each pericenter pass.  The fractional difference in pericenter distance between this and nearly total destruction of the star is thus small.  It is not impossible that the star really is balancing on this knife edge, and the uniqueness of HLX--1 makes it reasonable to contemplate such unlikely circumstances, but it is a priori improbable that the system would be in this state.  A possible solution is that the donor star is in the empty loss cone regime of two-body dynamics, in which the pericenter distance changes by a very small fraction of itself in every orbit.  

Another possibility worth considering is that the donor is a pulsating variable, so that it is the pulsation period rather than an orbital period which drives the outbursts.  The slight irregularity of such pulsations could explain the moderate ($\sim 2$ week) delay in the onset of the fifth outburst.  However, such models face serious difficulties related to the time scale.  The periods of the observed p and f modes are of the order of the dynamical time scale $(G\rho)^{-1/2}$ or {\it shorter} (see \citealt{2013ARA&A..51..353C} for a recent review of asteroseismology), which would mean that for the star to be overfilling its Roche lobe due to pulsations, its orbital period would have to be longer than the pulsation period.  This, in turn, would lead to disk inflow time scales that are much longer than the pulsation period.  These time scales involve diffusion, so it is difficult to see how any significant amplitude of variability could be maintained.  Gravity modes of pulsation can have periods longer than the dynamical period of a star \citep{2013ARA&A..51..353C}, but these are extremely weak and thus seem unlikely to produce the observed behavior.

Motivated by these challenges, we now explore a model in which the IMBH is fed by a wind.

\section{Wind accretion model}
\label{sec:wind}

The reason that previous modelers have disfavored wind accretion (e.g., \citealt{2011ApJ...735...89L}) is that the implied peak accretion rate $\sim 2\times 10^{-4}~M_\odot~{\rm yr}^{-1}$ is extremely high for a wind.  Indeed, such wind loss rates are at or above the very top end for Wolf-Rayet stars \citep{2000A&A...360..227N} and as we will see, in our preferred model only $\sim 1/4$ of the wind is captured at pericenter.  

In order to produce such a high wind rate, a special scenario is therefore required.  Motivated by the lack of ROSAT detections that we discussed in Section~\ref{sec:obs}, we propose that the IMBH in HLX--1 recently removed most of the envelope from a massive giant star.  See \citet{2012ApJ...757..134M,2013ApJ...777..133M} for recent discussions of tidal stripping of giants, although note that they focus on relatively weak tidal interactions whereas we hypothesize deep plunges in which most of the envelope is removed.  Massive giants are close to their critical luminosity; for example, according to runs with the StarTrack population synthesis code \citep{2002ApJ...572..407B,2008ApJS..174..223B}, after $\sim 27$~Myr a solar metallicity star with a zero-age main sequence mass of $10~M_\odot$ has a helium core mass of $2.8~M_\odot$ and a luminosity that would be $\sim 30$\% of the Eddington luminosity for that core mass. 

Thus if, as in the simulations of \citet{2013arXiv1307.6176B}, half or more of the total mass of the star were to be lost due to tidal disruption by the IMBH, and if the opacity is moderately enhanced above the Thomson scattering opacity, then because the luminosity of the star (which is generated in the core) would be unchanged but the mass would have decreased, the luminosity could well be super-critical and thus lead to an extremely high, if temporary, wind loss rate.  According to \citet{2013arXiv1307.6176B} not all of the envelope would be ejected.  Depending on the details of the wind, there could therefore also be a change in the wind loss rate with time after the partial disruption.  If most of the envelope is ejected dynamically and asymmetrically, the core and remaining envelope could be injected into a bound orbit.  This is the scenario we will consider; this would be a rare event, but again the uniqueness of HLX--1 suggests that uncommon scenarios can be considered.

In this section, we explore the consequences of the wind accretion model.  In Section~\ref{sec:windacc} we discuss the dynamics of wind capture.  Then, in Section~\ref{sec:windabs} we note that fast winds will populate the region over and around the disk with a significant excess of material.  We show that this is likely to generate an absorption-like feature in the $\sim 0.9-1.1$~keV range that is consistent with the data, although at low significance.

\subsection{Wind dynamics and accretion}
\label{sec:windacc}

The prime advantage of wind accretion compared to Roche lobe overflow is that if the wind speed is comparable to or larger than the orbital speed, then the net angular momentum of the captured matter can be much less than the orbital angular momentum.  Thus when the wind self-collides and cools, the resulting disk is small and hence the inspiral and decay times are as well.  Moreover, when the pericenter distance and wind speed are such that the time roughly matches what is seen in HLX--1, a slight increase in the wind speed reduces the peak flux by a few percent and more importantly reduces the decay time by the observed tens of percent.

To see this, suppose that a wind of speed $v_{\rm wind}$ emerges spherically from a star at the pericenter of its orbit, where it has a speed $v_{\rm peri}$ and where the escape speed from the IMBH at that distance is $v_{\rm esc}$.  Of the particles in the wind that are bound to the central IMBH, what is the average angular momentum compared to the angular momentum of a circular orbit at $r_{\rm peri}$?

To calculate this, we first normalize the other two speeds by $v_{\rm esc}$, and represent those speeds with hats over the $v$: ${\hat v}_{\rm wind}\equiv v_{\rm wind}/v_{\rm esc}$ and ${\hat v}_{\rm peri}\equiv v_{\rm peri}/v_{\rm esc}$.  We then set up a coordinate system in which the $+z$ axis, i.e., $\theta=0$, points in the direction of the orbit at pericenter and the center of the system is at the pericenter point.  The total speed of the wind relative to the IMBH is independent of the azimuthal angle, and is given by
\begin{equation}
{\hat v}^2_{\rm tot}=({\hat v}_{\rm peri}+{\hat v}_{\rm wind}\cos\theta)^2+{\hat v}_{\rm wind}^2\sin^2\theta\; .
\end{equation}
If ${\hat v}^2_{\rm tot}\geq 1$ the wind is at escape speed relative to the IMBH, and will therefore not accrete onto the hole.  Otherwise, it will accrete.  Thus to determine the average angular momentum of the accreting matter, we need to determine the boundary angle $\theta$ that divides escaping from retained material and then integrate over the retained $\theta$.  Writing $\mu\equiv\cos\theta$, we find

\begin{equation}
\mu_{\rm bound}={1-{\hat v}^2_{\rm peri}-{\hat v}^2_{\rm wind}\over{2{\hat v}_{\rm peri}{\hat v}_{\rm wind}}}\; .
\end{equation}
If $\mu_{\rm bound}>1$ in this equation, then all angles are allowed.  If $\mu_{\rm bound}<-1$, no angles are allowed.

The net average specific angular momentum is given by
\begin{equation}
\langle\ell\rangle/r_p={\int_0^{2\pi}d\phi\int_{-1}^{\mu_{\rm bound}}({\hat v}_{\rm peri}+{\hat v}_{\rm wind}\mu)d\mu\over{\int_0^{2\pi}d\phi\int_{-1}^{\mu_{\rm bound}}d\mu}}\; .
\end{equation}
Here $r_p$ is the pericenter distance.  Normalized to the specific angular momentum of a circular orbit at $r_p$, we find
\begin{equation}
{\langle\ell\rangle\over{\ell_{\rm circ,peri}}}=(1+e)^{1/2}-2^{-3/2}\left[(((1+e)/2)^{1/2}+{\hat v}_{\rm wind})^2-1\right]/((1+e)/2)^{1/2}\; ,
\end{equation}
where here we have used ${\hat v}_{\rm peri}=[(1+e)/2]^{1/2}$ for an orbit of eccentricity $e$.

As a specific example related to HLX--1, we assume as before that the mass transfer rate at pericenter is $2\times 10^{-4}~M_\odot$ and that $M=10^4~M_\odot$.  Suppose that in the first cycle the characteristic decay time, which we set equal to the inspiral time from the outer edge of the disk, is $10^7$~seconds.  If we also assume a high but plausible wind speed of $3000~{\rm km~s}^{-1}$ (see \citealt{2000A&A...360..227N} for some measured terminal wind speeds from Wolf-Rayet stars; note that for our postulated stripped giant, the wind speed could be even higher), then we find that for a pericenter distance $r_p=2.465$~AU and thus $e=0.8856$ the specific angular momentum corresponds to a circularization radius of 0.0633~AU, which from our previous calculation gives an inspiral time of $10^7$~seconds for $\alpha=0.1$.  A slight increase in wind speed, to $3025~{\rm km~s}^{-1}$, gives a circularization radius of 0.0517~AU and an inspiral time of $5\times 10^6$ seconds.  Note that these pericenters are roughly ten times the distance at which the star would start to transfer mass via Roche lobe overflow.

The fraction of the wind that is captured is just $(1+\mu_{\rm bound})/2$.  In our example this fraction is 0.226 when $v_{\rm wind}=3000~{\rm km~s}^{-1}$ and 0.224 when $v_{\rm wind}=3025~{\rm km~s}^{-1}$.  Thus if the wind loss rate is constant, the accretion rate at pericenter drops by just a percent.  This also means that if at peak the accretion rate is $\sim 2\times 10^{-4}~M_\odot$~yr$^{-1}$, the mass outflow rate needs to be a few times larger than this.  That rate would exceed any known wind rate \citep{2000A&A...360..227N} by a factor of a few.  This is one reason that we consider the possibility that the IMBH recently stripped most of the envelope of a massive giant; for a few years, the wind loss rate from the remaining core and envelope could be much higher than is normally possible.  Such a scenario could also explain the rapidly decreasing integrated luminosity of the outbursts and the non-detection in the ROSAT observations.

The specific numbers here should be considered only illustrative, because it is unfortunately not straightforward to go from our wind capture model to the expected flux as a function of time.  One reason is that much of the wind that is eventually captured is only marginally bound, and hence will return to the disk over a time scale that could be significantly longer than the pericenter passage time.  Another complication is that the disk structure may in fact not be similar to the Shakura-Sunyaev equilibrium structure, given that the matter comes in pulses rather than steadily.  Both of these issues will also apply to Roche lobe overflow models.  An additional complexity unique to the wind accretion model is that although most of the accreted matter will be donated near pericenter passage, some will arrive before or after pericenter, and because this matter comes from parts of the orbit that have smaller escape speeds than at pericenter, the specific angular momentum will be less and could even be negative.  Thus the actual time variation of the structure of the disk as well as the accretion rate and resultant flux will actually have more complex dependencies on the orbit than we have used in our simple model.  Some of this complexity could explain the observed few-week delay of the onset of the most recent outburst.

\subsection{Extra absorption induced by a wind}
\label{sec:windabs}

In the Roche lobe overflow scenario, the donated matter will be largely confined to a disk.  In contrast, in our wind scenario a significant amount of the matter will be distributed throughout the volume.  More specifically, suppose that the wind rate from the donor is $10^{-3}~M_\odot~{\rm yr}^{-1}$, and that the wind speed is 3000~km~s$^{-1}$.  At this speed, the wind would cross the $\sim 20$~AU semimajor axis of the donor's orbit in $\approx 10^6$ seconds, so if the matter is distributed uniformly in a sphere the number density would be
\begin{equation}
n\approx 10^{-3}~M_\odot~{\rm yr}^{-1}\times 10^{57}~{\rm nucleons}~M_\odot^{-1}\times 10^6~{\rm s}/[4\pi/3(20~{\rm AU})^3]\approx 3\times 10^8~{\rm cm}^{-3}\; .
\end{equation}
This implies a column depth of $3\times 10^8\times 20~{\rm AU}\sim 10^{23}~{\rm cm}^{-2}$, which is far in excess of the $N_H=3\times 10^{20}$ that was inferred from X-ray fits.

We were therefore motivated to look for spectral signatures of this excess column.  Given the strong photoionizing flux on the wind, we take a typical set of parameters and run XSTAR simulations to determine whether we would expect detectable absorption features.

\begin{figure}[htb]
\begin{center}
\plotone{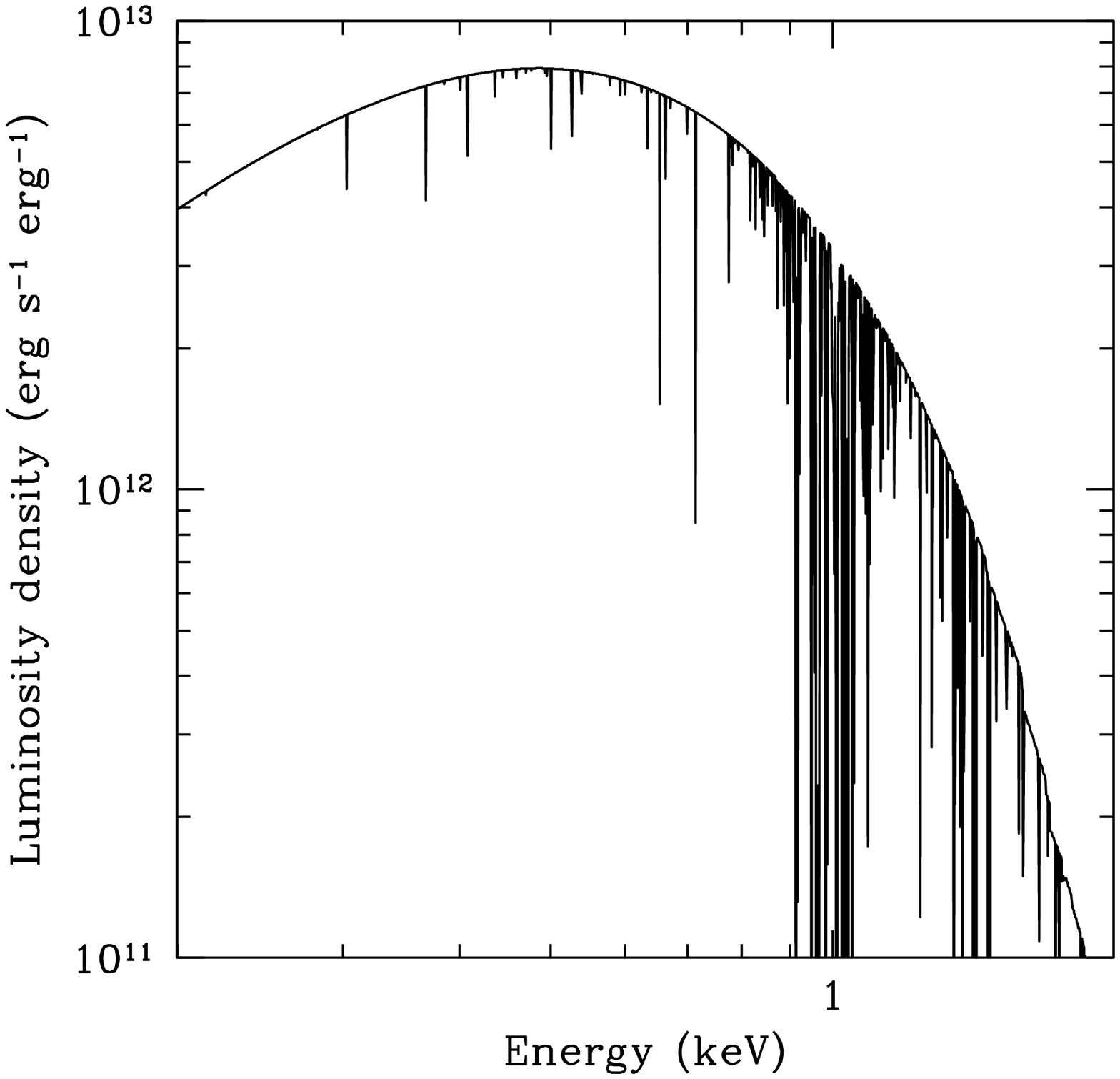}
\caption{A plot of the luminosity density versus energy for the XSTAR simulation of a $2\times10^6$ K blackbody passing through a strong stellar wind.  The dips that are deviations from a smooth curve indicate where the absorption from the photoionized material is strongest.  We note in particular that from about 0.9-1.1 keV, about 30\% of the flux is absorbed.  As we see in Figure~\ref{fig:spectrum}, this is consistent with what is observed from HLX--1.}
\label{fig:abs}
\end{center}
\end{figure}

\begin{figure}[htb]
\begin{center}
\plotone{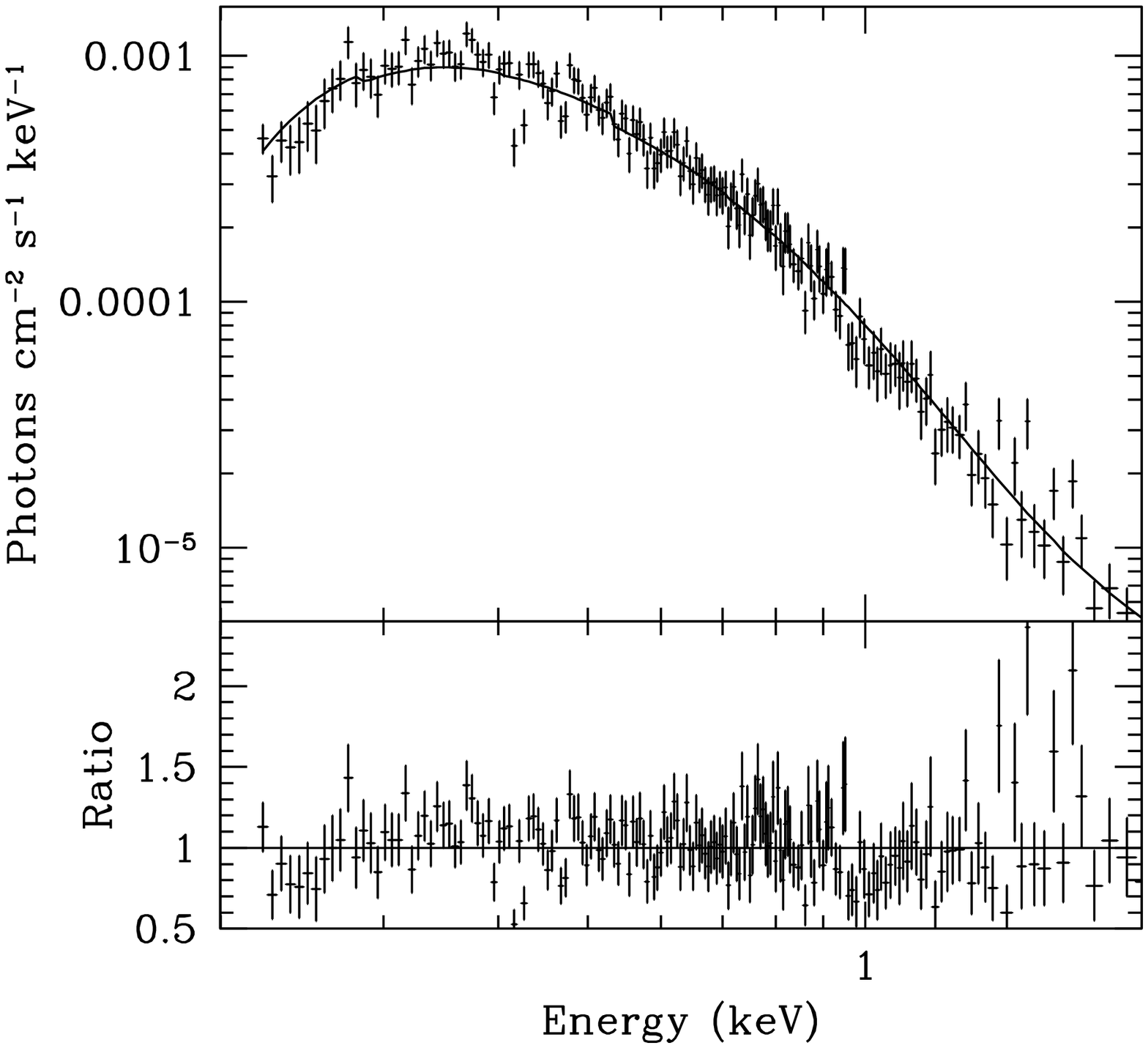}
\caption{XMM-Newton pn spectrum in the 0.2--2~keV range of HLX--1 taken on 28 November 2008 (OBSID=0560180901), fitted with an absorbed irradiated disk model (tbabs*diskir in XSPEC).  This observation was selected as it is the highest signal to noise X-ray spectrum taken of HLX--1.  We draw attention to the region around 0.9--1.1~keV, which (as predicted in our simple wind model) has a possible spectral deficit of a few tens of percent.}
\label{fig:spectrum}
\end{center}
\end{figure}

The parameters we used in the XSTAR calculation are: covering fraction 1.0, wind temperature $10^5$ K, number density $4\times10^8$ cm$^{-3}$, column density $10^{23}$ cm$^{-2}$, ionization parameter $2\times 10^4$ (appropriate for a $2\times 10^6$~K blackbody with a total luminosity of $10^{42}$~erg~s$^{-1}$ in the $\sim 20$~AU radius volume), and solar abundances.  The absorption from the ionized wind is negligible over most of the energy range for which the X-ray spectra have a substantial signal to noise ratio (see figure \ref{fig:abs}).  However, from about 0.9-1.1 keV, there is a series of significant absorption lines from the ionized material.  Running a boxcar filter over the spectrum with a width of 0.1 keV, we find that in this energy range, about 30\% of the flux is absorbed.  As can be seen in Figure~\ref{fig:spectrum}, this is consistent with what is seen from HLX--1.  We emphasize, however, that although the detection of the predicted signal is encouraging for the wind scenario, this should not be interpreted as conclusive verification of the model.  Our input parameters and geometry are only notional, and indeed we simply lack the information required to produce a more definitive model.  Nonetheless, the potential excess absorption is a point in favor of a wind model rather than a Roche lobe overflow model.

\section{Dynamics}
\label{sec:dynamics}

The presence of a young massive cluster around the IMBH \citep{2012ApJ...747L..13F,2014MNRAS.437.1208F} suggests that we should investigate whether dynamical interactions with other stars could alter the orbit significantly over the few years during which we have observed the system.  As we will now discuss, we find that this is highly unlikely for realistic densities.

The mass and age of the cluster surrounding HLX--1 has been debated in the literature.  The most recent analysis \citep{2014MNRAS.437.1208F} finds strengthened evidence for a young cluster with an age of $\sim 20$~Myr that might have been produced during a merger event with ESO~243--49.  The best fit mass of the young component is $9\times 10^4~M_\odot$, but within the substantial uncertainties in the luminosity, age, and metallicity of this component \citet{2014MNRAS.437.1208F} conclude that the cluster could have a mass between $\sim 5\times 10^2$ and $\sim 6\times 10^6~M_\odot$.  \citet{2014MNRAS.437.1208F} do not find any positive evidence for an older stellar component, but their weak upper limit of $\sim 10^{39}$~erg~s$^{-1}$ from such a component implies a fairly nonrestrictive upper limit of $\sim 10^6~M_\odot$ on the mass of a $\sim 10$~Gyr old population.

Our dynamical inputs are therefore highly uncertain.  Suppose for the sake of argument that we assume a total mass of $10^6~M_\odot$ and suppose that the relation $M/10^8~M_\odot\approx (\sigma/200~{\rm km~s}^{-1})^4$ that holds for supermassive black holes (e.g., \citealt{2009ApJ...698..198G}) also holds for the IMBH in HLX--1.  Then the radius of influence is $r_{\rm infl}=2GM/\sigma^2\approx 4\times 10^4~{\rm AU}$ for this system.  If the number density within the radius of influence scales as $n\sim r^{-2}$, then the total mass of stars within the semimajor axis $a$ is $\sim (a/r_{\rm infl})M$, or $\sim 5-10~M_\odot$.  A single stellar-mass black hole would thus mean $m=10~M_\odot$ and $N=1$, whereas if the perturbers are solar-type stars then $m=1~M_\odot$ and $N={\rm few}$.  In this circumstance, the change in specific angular momentum over a time $t$ by resonant relaxation is a few times $10^{-4}$ times the circular angular momentum, per orbit (see the expressions in \citealt{1996NewA....1..149R}).  This is roughly consistent with what would be needed to explain the observed changes in the decay time.

The problem with this argument is that the required stellar number densities are enormous: a few stars within $\sim 20~{\rm AU}\approx 10^{-4}$~pc implies $n\sim {\rm several}\times 10^{12}$~pc$^{-3}$.  Thus the expected time for direct collisions between solar-type stars ($R\approx 10^{11}$~cm and therefore $\Sigma\sim 3\times 10^{22}$~cm$^2$) at the orbital speeds $v=\sqrt{G10^4~M_\odot/20~{\rm AU}}\approx 700$~km~s$^{-1}$ is $\tau=1/(n\Sigma v)\approx 2\times 10^6$~yr.  This is significantly shorter than the age of even the young component of the cluster, so we expect stars to have undergone multiple collisions.  In addition, it is difficult to see how the cluster would have evolved to such a centrally compact state even if the stars were effectively point masses.  If the cluster does have a velocity dispersion of $\sim 20$~km~s$^{-1}$ as suggested by the $M-\sigma$ relation, then binary heating should be able to hold off deep core collapse \citep{2012ApJ...755...81M}, and the IMBH itself could be an even more efficient heat source \citep{2008ApJ...686..303G,2013MNRAS.435.3272T}.  Thus we expect the number density to be orders of magnitude less than would be necessary for significant dynamic evolution in a few years.  This is why we suggest that changes in the wind outflow rate and possible speed are more likely to cause the changes seen from outburst to outburst.

For completeness we also address the possibility that the system is in a Kozai resonance state.  In such a state, the binary formed by the donor and the IMBH has long-term interactions with a tertiary star, in such a way that over many orbits the eccentricity and inclination of the inner binary orbit oscillate while keeping the binary semimajor axis constant.  In this model, the $\sim 1$~yr period we see is the time needed for a full cycle in the eccentricity of the inner orbit, i.e., 1~yr is many inner orbital periods.  There are several problems with this model, one of which is that when the peak eccentricity over a Kozai cycle is moderate to high, one would expect significant modulation of the accretion rate on the binary orbital period.  Given that, for an IMBH mass $M$, a tertiary mass $m_3$, a tertiary semiminor axis $b_3$, and a binary semimajor axis $a$ the Kozai cycle timescale is $\sim (M/m_3)(b_3^3/a^3)P_{\rm orb}$ (e.g., equation (8) of \citealt{1997AJ....113.1915I}), the binary orbital period $P_{\rm orb}$ could be a few thousand seconds or less.  Thus we would expect modulation on this timescale, which is not seen.  In addition, this would imply a tidal radius of $\ltorder 0.01$~AU, which would suggest a decay time that is much {\it less} than is seen.  We therefore do not favor a Kozai resonance explanation for HLX--1.

Finally, we can return to our proposed scenario of tidal disruption of the envelope of a giant star.  From \citet{1988Natur.333..523R}, the most bound material in the ejecta has a semimajor axis that is $\sim (M/m)^{1/3}$ times the tidal radius.  Thus the time for the first bound matter to return to the IMBH will be $\sim (M/m)^{1/2}$ times the orbital period at the tidal radius.  Given that the giant has a substantial density gradient, we could take the pericenter of the current orbit as the tidal radius; this corresponds to having the bound matter originate from the densest part of the envelope that is stripped from the star.  Thus if $M/m\sim 10^4$, the time for the first matter to return will be $\sim 10^2$ times the orbital time at the pericenter, which in turn is $(1-e)^{3/2}$ times the $\sim 1$~yr current period of the binary.  If $e\sim 0.8-0.9$ this suggests that the first return time is of the order of a decade to decades.  Thus several solar masses will return over that period.  This rate is very super-Eddington, which suggests that in tens of years the source will become persistently bright, and will remain super-Eddington ($\gtorder 10^{-4}~M_\odot~{\rm yr}^{-1}$) for centuries.

\section{Conclusions}
\label{sec:concl}

The magnitude and evolution of the decay time seen from the outbursts of HLX--1 are difficult to explain in a Roche lobe overflow model.  This motivates study of a wind accretion model, which as we have shown can naturally explain the observed decay times if the wind speed evolves slightly from outburst to outburst.  The price to pay for the wind model is steep: the required accretion rate is near the top end of what is seen from Wolf-Rayet stars, and possibly a factor of a few higher yet.  We suggest that the needed wind loss rate could have been provided by a recent tidal disruption of a massive red giant, which left behind a helium core with a small hydrogen envelope.  We also note that the wind model would imply that there is considerable extra absorption in the system; motivated by this prediction, we searched for and found tentative evidence of such absorption.  This provides encouraging but not definitive support for our model.  On the long term, we expect that the source will become persistently bright within roughly a decade to decades due to the accretion of returning bound matter.  Further monitoring of this source will be crucial to determine its nature.

\acknowledgements

This work was supported in part by a grant from the Simons Foundation (grant number 230349 to MCM) and NASA ATFP grant NNX12AG29G.  SAF is the recipient of an Australian Research Council Postdoctoral Fellowship, funded by grant DP110102889.  We thank Tal Alexander, Chris Belczynski, Tamara Bogdanovi\'c, Jean-Pierre Lasota, Elena Rossi and Natalie Webb for helpful discussions. MCM also thanks the Department of Physics and Astronomy at Johns Hopkins University for their hospitality during his sabbatical.  All the authors thank the {\it Swift} team for the monitoring campaign that has revealed so much about HLX--1.

\bibliography{ms}

\end{document}